# OOP and Its Calculated Measures in Programming Interactivity

Onu Fergus U[1], Osagie S. U. M.[2], John-Otumu M. A.[2], Igboke M. E.[3]

*[1, 3]Department of Computer Science, Ebonyi State University, Abakaliki, Nigeria*
*[2]Department of Mathematics and Computer Science, Benson Idahosa University, Benin, Nigeria*
*[2]Department of Computer Science, Ambrose Alli University, Ekpoma, Nigeria*

***Abstract:*** *This study examines the object oriented programming (OOP) and its calculated measures in programming interactivity in Nigeria. It focused on the existing programming languages used by programmers and examines the need for integrating programming interactivity with OOP. A survey was conducted to measure interactivity amongst professionals using certain parameters like flexibility, interactivity, speed, interoperability, scalability, dynamism, and solving real life problems. Data was gathered using questionnaire, and analysis was carried out using frequency, percentage ratio, and mean in arriving at a more proactive stand. The results revealed that the some of the parameters used are highly in support of the programming interactivity with OOP.*

**Corresponding address:** macgregor.otumu@gmail.com

## I. Introduction

Since the introduction of computer to human society it has bridged the divided world of information and thereby given back interest on investment (return on investment) in business field / profession. So, having the understanding on how object interacts and share resources between interfaces will go a long way in helping researchers, software engineers and young professional programmers in the software industry gain first hand information on how current programming languages uses objects of different classes to execute its operations.

Objects provide an ideal mechanism for implementing abstract data types, which includes stacks, trees, and hash tables. Objects can also be seen to be a nice implementation vehicle for abstract data types since the data stored by the objects can represent the abstract data types and the object's interface can represents the abstract data type's set of operations.

One of the original purposes for object languages was to model applications that have multiple objects that may be operating simultaneously. Object oriented languages uses multi and distributed threading system in building control within applications developed as compared to conventional imperative languages like C having problems modelling complex applications because they have a single thread of control. Object languages thereby solve real life problems by making everything an object and having control reside within each object, i.e. at any given moment multiple objects could be executing an operation and communicating with other objects with the concept of message passing; which is simply an invocation of an operation in another object.

### 1.1 Research Objectives

The overall objective of this research work is to carry out a study on Object Oriented Programming (OOP) and its calculated measure in programming interactivity. The specific objectives are as follows:

1. To review the concept of object and it role in programming interactivity.
2. To know the possibility of dissimilar interface with different classes of objects interoperate in an OOP environment.
3. To measure and ascertain OOP flexibility amongst professionals in the field of computing.

## II. Related Literature

### 2.1 Definition and Evolution Object-Oriented Programming (OOP) Concept

Many authors in different literatures have defined object-oriented programming in distinct manner leaving readers to figure out the actual meaning of OOP by exploiting the laid down objects structure. Object oriented program is a program that consists of one or more objects that interact with one another to solve a problem.[1]

Object Oriented Programming (OOP) is an approach to program organization and development that attempts to eliminate some of the pitfalls of conventional programming methods by incorporating the best of structured programming features with several powerful new concepts [2]. It is a new way of organizing and developing programs and has nothing to do with any particular language. However, not all languages are suitable to implement the OOP concepts easily.

This research paper defines object-oriented programming from a different perspective and how it relates to everyday programming. "Object-Oriented Programming can be define as act of programming which involves the reuse of entity (object) in modelling new and real shape of realistic nature with "identity" that enable programming specifications and procedure, "State"(which determine the object attribute) and "Behaviour" (which determine the interactivity) within objects of different classes".

**Table 1: Evolution of Object Oriented Programming Languages**

| S/N | OOP Languages | YEAR | DECRIPTION |
|-----|---------------|------|------------|
| 1. | Simula | 1967 | Developed for mainly simulating discrete invent systems. |
| 2. | Smalltalk | 1970s | Developed for simulation and graphics-oriented systems |
| 3. | Ada | 1980 | A structured, statically typed, imperative, wide-spectrum, and object oriented high-level computer programming language built to support extremely strong typing, explicit concurrency, offering tasks, synchronous message passing, protected objects and non-determinism |
| 3. | C++ | Early 1980s | First object oriented language to widely used commercially. |
| 4. | Java | Early 1990s | A simpler version of C++ developed by Sun Java. It is meant to be a programming language for video-on-demand applications and Internet applications development. |

A closer look at the evolution of object oriented programming languages brings to the understanding of its concepts and usefulness in the world of programming. Today new generation of programming languages utilizes the coherent principle of object exchange in achieving its goal. [3, 4, 5, 6]

## 2.2 Features of OOP
Object Oriented Programming and Languages have a well robust feature that gives adequate support to interfaces interactivity and these include the following:

**Object:** The findings have revealed that object has three main characteristics that translated to its flexible nature which includes:
i. State
ii. Behaviour
iii. Identity

The entire concepts above shows that object are tangible entities. Figure 1(a) shows the state of the ball (the ball is at the middle of the pitch) awaiting other objects (entities) such as figure1(b) for interactions, thereby representing the behaviour of different entities having a well define name figure1(c) that represents the principle to which object function in current programming languages. The flexibility of this approach enable interface to communicate freely. Every component Interact with each other in creating a suitable object oriented program. [7]

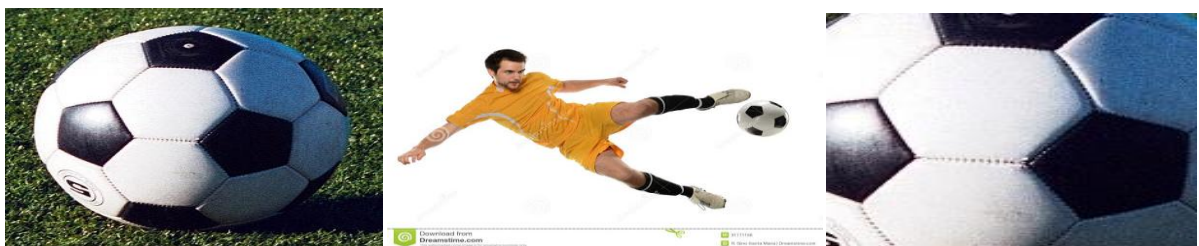

(a) "State" (On the soccer field)   (b) "Behaviour" (Entities interactions)   (c) "Identity" (Soccer ball 124)
**Figure 1:** Illustrating the three main characteristics of Object [7]

**Classes:** As discussed briefly above classes are group of objects with the same characteristic and they share common properties. For easy interactivity, objects are built into classes and the classes interact within a given set of definitions. A class has three kinds of member [8]
1. Field that specify the kind of data that the objects hold;
2. Methods that specify the operations that the objects can perform;
3. Constructors that specify how the objects are to be created.

**Modularity:** Object oriented programming help to create a well modular structure in interface interactivity by providing interface to which other units of code can be written and allow access without copying code or writing explicit code to discover it implementation details. It is an interface with sets of share names for procedure or object. Module has the capability of running flexibly in a separate machine of different entities

**Reusability:** This is one of the benefits of object oriented concept. It uses existing code to implement new function/procedure. This eliminates the fundamental problem of the early programming languages; thereafter create a flexible approach towards software development.

**Mechanisms of Abstraction:** The world is made up of objects and they are of different manners, these objects are grouped into classes for proper identification. Abstraction is simply the process of selecting the significant attribute for easy classification of a given set of objects.

**Encapsulation:** the design structure of object oriented concept has a mechanism to which objects/data are hidden from the outside world.

**Polymorphism:** Object oriented programming / languages foundation is in object class function. Every object has its characteristics and in programming they must be able to use such attribute without interference. The ability of the object oriented programming languages to process each object in a separately form (differently) by treating each object data type, classes and redefining methods for classes derivations is known as polymorphism

**Inheritance:** Many had it that a good parent must leave possessions for his/her children and their children also doing same to their next generation. As it is popularly know it is the transfer of one wealth (properties), attributes (trait) or feature from one generation to another generation say parent to child. In object oriented programming object or class passes some attributes to subclass. This can be seen from supper-class and subclass relationship where setting features by the supper-class is passed on to the subclass. Rather than building at fresh inheritance utilises the reuse mechanism in implementing object oriented programming project. This is the principle to which current programming languages are built.

### 2.3 Benefits of Object Oriented Programming (OOP)

OOP offers several benefits to both the program designer and the user. Object orientation contributes to the solution of many problems associated with the development and quality of software products. The new technology promises greater programmer productivity, better quality of software and lesser maintenance cost.

The following are some of the major benefits of OOP identified:
1. Through inheritance, redundant codes can be eliminated.
2. Programs can be built from standard working modules that communicates with one another, rather than having to start writing the code from the scratch. This leads to saving development time and higher productivity.
3. The principle of data hiding helps the programmer to build secure programs that can not be invaded by code in other parts of a program.
4. It is possible to have multiple instances of an object to co-exist without any interference.
5. It is possible to map object in the problem domain to those in the program.
6. It is easy to partition the work in a project based on objects.
7. The data-centered design approach enables us to capture more detail of a model.
8. Object oriented system can be easily upgraded from small to large system.
9. Message passing techniques for communication between objects makes interface descriptions with external systems much simpler
10. Software complexity can be easily managed.

### 2.4 Application Interface Development Issues

A program design without interface is as good as nothing. Computer program designed utilises interface to communicate with other programs either for resource sharing or data sharing. There are many reasons why a program cannot be designed without interface but for the purpose of this study we will focus on key issues such as lack of openness, connection between the underlying applications and lack of support beyond the coding phase [9],[10]. Also we will examine the security issues. There are several issues facing the interface and according flexibility. Creating good user interface for system is a difficult task. User interface software is often large, complex, and difficult to debug and modify. It often represents a significant fraction of the code, frequently ranging from 40 to 60 percent. Good interfaces he said are easy to use frequently require several cycles of designing, development, testing, and refining. Consequently, better tools are needed for all aspects of user interface development, ranging from support of complex programs to rapid prototyping.

Another big issue facing interface development is the concern for security. In designing or building interfaces, programmers are more concerned over the manner at which hackers or invaders manipulates interface in order to gain access to valuables. Also, OOP provides a systematic approach to solving this problem. As discussed above, every object is encapsulated in such a way that they act as a **FIREWALL** between outside world and the object itself. Therefore making it easier for classes to call object using **methods** and **messages** approach without been known to the outside world.

## 2.5 Dynamics of Applications Interfaces Interactivity

The first generation programming languages application interface suffers major setback as a result of system engender, this is similar to computer design by different vendors in the early stage of computer productions. Communications between these computers (from different vendors) were impossible due to compatibility and interoperability issues. This period was regarded as "**systems engender**".

In interface interactivity OOP concept /model has bridge the divided world of interface working as a single entity. Current programming has shown that interface written in Java, VB, C++ or Hypertext Mark-up language can interact freely and this is as a result of the dynamism (scripts exchange) to which object oriented programming languages were built.

## 2.6 Interface Interactivity and the OOP Concept

Interface is component of software or program which gives the user access to use the application. To this end, we can say that interface is the entrance door to which programmer or computer users gain access to other information available in a given computer. Every system has interface and to structured program, it is the part where authentication of a given user is required. It is the first part of a machine when switched on to ask for some valid data. Interface interactivity and OOP concept are one and the same. The possibility for interface to interact with another interface centred on the laid down concept of OOP.

## 2.7 Implementing Interfaces with OOP

Modern day programming languages use object oriented concept in solving real life problems. OOP creates opportunity to build programs from existing component. According to [11] building a graphical user interface (GUI) in Java, the JDK has a GUI component to which interface can be built, the JDK which means Java development toolkit is in integration with the integrated development environment (IDE). This library enables programmer to pick component to build new interface rather than building from the scratch.

There are two basic types of Java graphical user interface and these include:
➢ Applet
➢ Standalone graphical user interface application

For the purpose of this research work, we shall examine the one of the applet using Java programming language Applet like any other program does not run directly on the platform it was created, rather it runs from different entity in initiating it operations. Applet program is not too large but has the sole capability of running via a browser acting as the programming housing the program.

Applet runs well from the following browsers:
➢ Mozilla Firefox
➢ Internet Explorer
➢ Opera Mini
➢ Google Chrome etc

Irrespective of how applet is examined, its class has extension of the java.applet.applet class. There are more varieties of functions with constant increase in java applet class.

## 2.8 Object Oriented Programming Scalability

Programmers that have spent years and time in working with different programming languages would attest to the fact that since object oriented concept introduction to programming languages, there has been tremendous improvement in programming. OOP is not just growing or making programming more flexible but expanding in an exponential form and its scalability has brought about flexibility, low cost, stress free, and return on investment to the software industries. It has also made life much more meaningful to the computer end users.

## III. Methodology

The following approaches were used in this research work in order to measure Object Oriented Programming (OOP) interactivity.

### 3.1 Data Collection

In order to obtain data from professionals in the field of computer science / information technology in academia (lecturers / programmers / Analyst / Students ), software practitioners in the software industry of Nigeria, and independent researchers in the field of computer science / IT, a questionnaire was designed and administered. The questionnaire was divided into three sections. The first section contains questions on socio-economical characteristics of the respondent. The second section contains questions regarding information of the respondent's basic skills, while the third section was designed to collect data about different characteristics statements based on the degree of acceptance or rejection to the point stated.

### 3.2 Data Analysis and Results Presentation

As a result of this survey, 100 copies of the questionnaire were administered, and 60 copies were received. The researchers used different methods to analyse the data gathered from the various sections of the questionnaire. The frequency and percentage ratio methods were used in sections A and B to determine the number of occurrence and percentage involved, while in section C, the mean was used to determine the level of support as it relates to OOP and its calculated measure in programming interactivity.

**Table 2: Socio-economic characteristics of respondents**

|  | Frequency | Percentage |
|---|---|---|
| (a)        GENDER |  |  |
| Male | 40 | 66.7 |
| Female | 20 | 33.3 |
| (b)        OCCUPATION |  |  |
| Researchers in I.T. field | 13 | 21.7 |
| Lecturers in Computer Sc. / IT | 26 | 43.3 |
| Software Practitioner | 3 | 5.0 |
| Programmer | 5 | 8.3 |
| Analyst | 3 | 5.0 |
| Student | 10 | 16.7 |
| (c)        EXPERIENCE IN PROGRAMMING |  |  |
| < 5 years | 29 | 48.3 |
| < 10 years | 19 | 31.7 |
| > 10 years | 12 | 20.0 |

Table 2 shows the distribution of the respondents' gender, occupation and experience in programming.

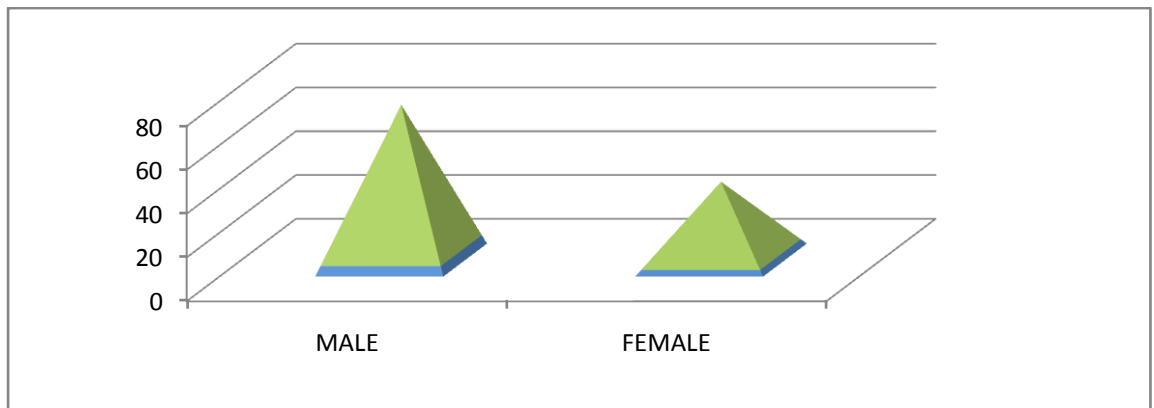

**Figure 2** Gender

In figure 2 (Gender), the male respondents have a total number of 40 with a percentage of 66.7%, while the female respondents have a total number of 20 with a percentage of 33.3%, i.e. the male are dominating in programming.

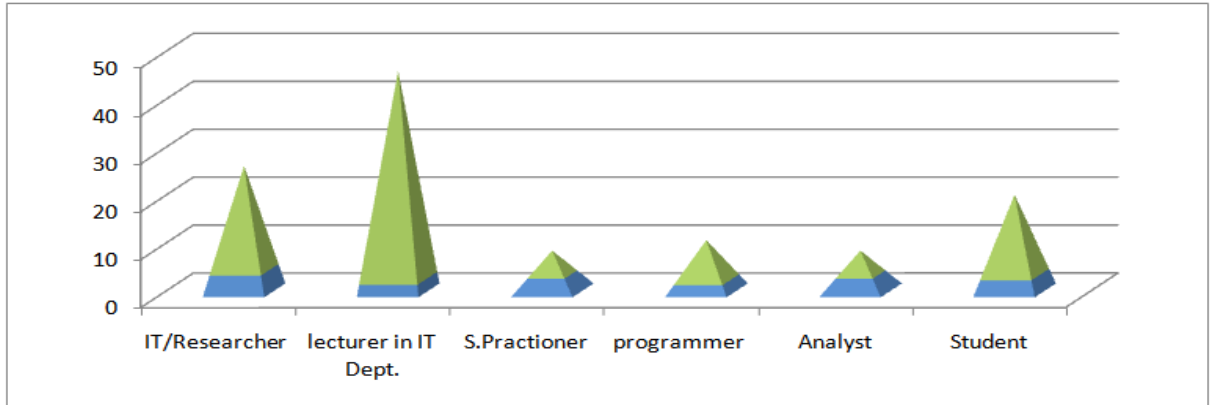

**Figure 3** Occupation

Under occupation the results in figure 3 shows that researchers in information technology field were 13 (21.7%), lecturers in the department of computer science/information technology were 26 (43.3%), industry software practitioners were 3 (5%), programmers were 5 (8.3%), Analyst were 3 (5%), and students were 10 (16.7%), again, this shows that the lecturers in computer science / information technology are strongly in support of the view.

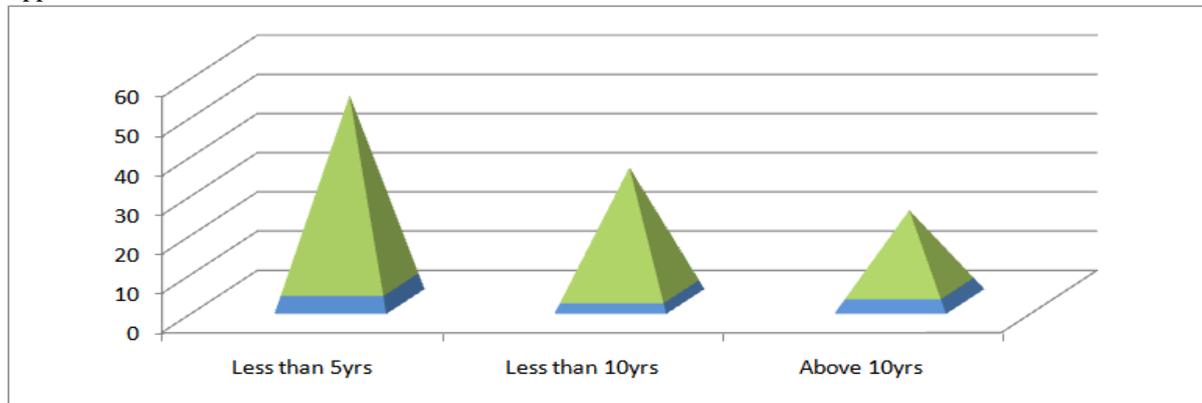

**Figure 4** Programming Experience

In terms of programming experiences, Figure 4 above also shows that 29(48.3%) respondents out of the 60 has less than 5 years coding experience, 19 (31.7%) respondents has less than 10 years experience, 12 (20%) respondents has over 10 years coding experience. This again shows that young programmers with less than 5 years programming experience are involved in OOP usage and strongly in support of it interactivity.

**Table 3:Distribution of respondents according to their coding, program methods and OOP skills**

|  | Frequency | Percentage |
|---|---|---|
| (a)       DO YOU CODE PROGRAMS | | |
| Yes | 42 | 70.0 |
| No | 9 | 15.0 |
| Not Exactly | 9 | 15.0 |
| (b)       PROGRAMMING METHOD PREFERRED | | |
| Procedural | 8 | 13.3 |
| OOP | 40 | 66.7 |
| Others | 12 | 20.0 |
| (c)       ARE YOU SKILFUL IN TERMS OF OOP | | |
| Yes | 29 | 48.4 |
| No | 8 | 13.3 |
| Not Exactly | 23 | 38.3 |
| (d)       LEVEL OF SKILFULNESS IN OOP | | |
| Highly Skilful | 5 | 8.3 |
| Moderately Skilful | 35 | 58.3 |
| Not Skilful | 20 | 33.3 |

Table 3 shows distribution of respondents according to their coding, program methods and OOP skilfulness.

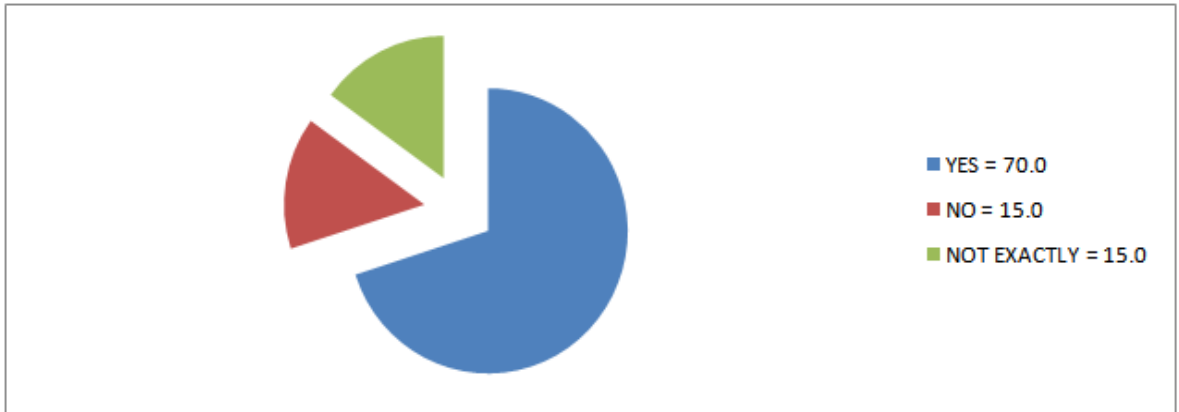

**Figure 5** Do you code programs?

With respect to Figure 5 (Program coding), 40(70%) of the respondents can code or write program, 9(15%) respondents can't code, while the other 9(15%) of the respondents are not so sure.

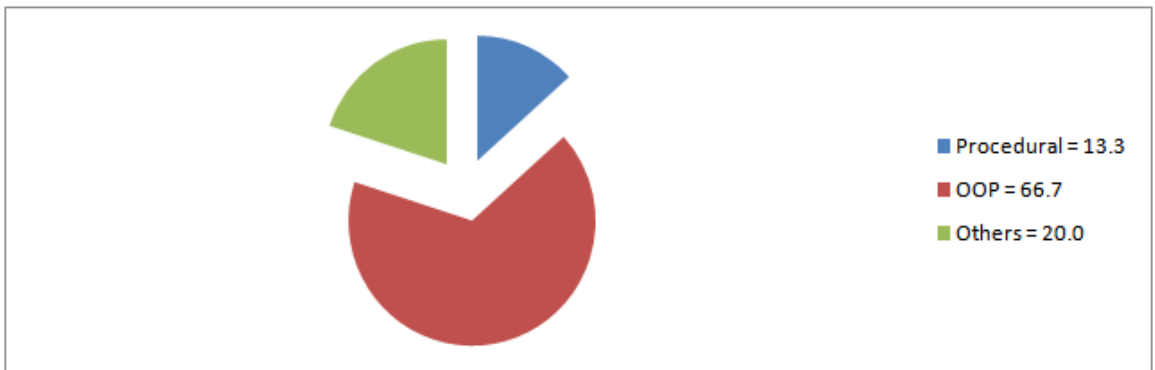

**Figure 6**  Programming method Preferred

In figure 6 (Preferred programming method), 8(13.3%) respondents out of the 60 respondents preferred the procedural method for programming, 40(66%) of the respondents preferred object oriented programming (OOP) approach, while 12(20%) of the respondents preferred other methods.

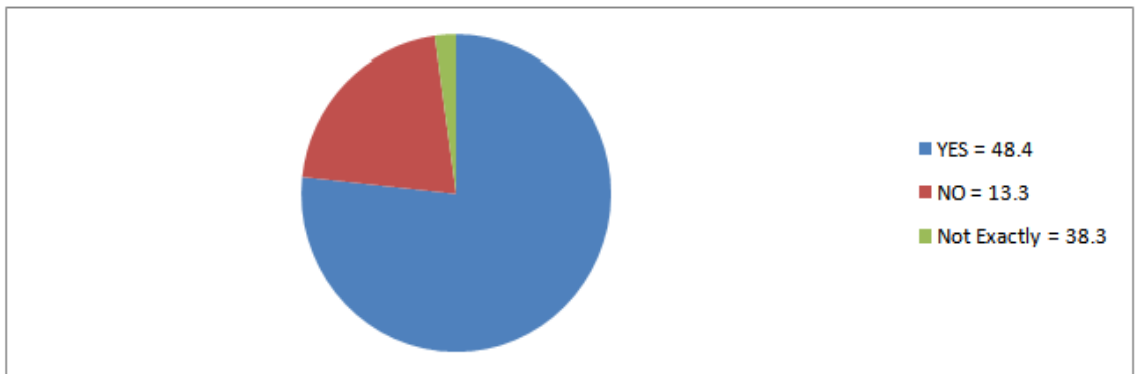

**Figure 7** Are you skilful in terms of OOP

In choice of OOP skills as shown in figure 7, pie chart, 28(46.7%) respondents claimed to be skilful in terms of OOP, 8(13.3%) respondents claimed not skilful at all, 23(38.3%) respondents are not certain, while 1 (1.7%) respondent did not respond to this question.

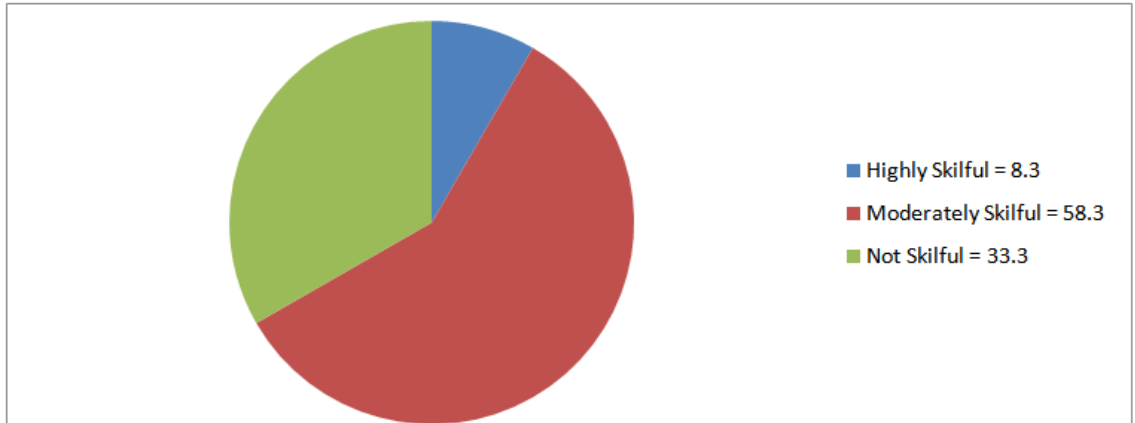

**Figure 8** Level of Skilfulness in OOP

Figure 8 shows the level of skilfulness with OOP designs. 5(8.3%) respondents claimed to be highly skilful in terms of OOP designs, 35(58.3%) respondents claimed to be moderately skilful, while 20(33.3%) claimed not to be skilful at all. In summary, there are more skilful programmers and they prefer using the OOP methods because of its flexibility and interactivity.

**Table 4: Object Oriented Programming Usage Experience of Respondents**

| S/N | Statements to Measure Object Oriented Programming (OOP) Interactivity | Mean | Standard Deviation |
|---|---|---|---|
| 1 | Introduction of OOP has made program design flexible for usage | 4.58 | .561 |
| 2 | OOP is highly interactive | 4.23 | .621 |
| 3 | OOP interactivity can be measured in term of objects design | 4.15 | .606 |
| 4 | OOP usage in program design is very fast as compared to procedural | 3.95 | .746 |
| 5 | Object design and usage in different OOP languages can interoperate | 3.80 | .879 |
| 6 | OOP is a scalable programming language | 4.13 | .676 |
| 7 | OOP scalability can be measured in terms of its improvement of the years | 3.73 | 1.071 |
| 8 | OOP can be used to implement interfaces which can solve real problems | 4.45 | .649 |
| 9 | Application interface interactivity written in OOP is more dynamic | 4.08 | .809 |
| 10 | OOP concept helps application development to be flexible and interactive. | 4.63 | .581 |
| | **\*Agreed (Mean > 3.00)** | | |

Table 4 shows the agreed mean, mean, and standard deviation values gotten as a result of the statistical analysis ran on data gathered from this section.

$$\text{mean } (x) = \frac{\sum fx}{\sum f(n)} \text{ that is, where } n_1 + n_2 + \ldots + n_n = n$$

The agreed mean value of 3.00 is the benchmark value for basic assessment of each statement computed mean value in table 4. The computed mean value for each statement as analyzed in Table 4 is compared against the agreed mean value for support of its measure of OOP interactivity. If the computed mean value is highly above the agreed mean value, then the statement is strongly in support of OOP interactivity. The statement is moderately or slightly in support of the argument statement, if the computed mean value is slightly above or in the same range with the agreed mean value or not in support at all, if the computed mean value of the statement is less than the agreed mean value.

## IV. Discussion

Table 4 presented OOPs calculated measures in programming interactivity. We observe the following:

i. **Introduction of OOP has made program design flexible for usage:** the computed mean value = 4.58. The mean value (4.58) is far above the agreed mean value which is 3.00 this shows that the statement is strongly in support of OOP and its calculated measures in programming interactivity.

ii. **OOP is highly interactive:** the computed mean value = 4.23. The mean value (4.23) is far above the agreed mean value which is 3.00 this shows that the statement is in agreement with OOP and its calculated measures in programming interactivity.

iii. **OOP interactivity can be measured in terms of objects designed:** the computed mean value = 4.15. The mean value (4.15) is also above the agreed mean value of 3.00. This shows that the statement is also in support of OOP and its calculated measures in programming interactivity.

iv.  **OOP is a scalable programming language:** the computed mean value = 4.13. The mean value (4.13) is above the agreed mean value which is 3.00 this again shows that the statement is in support of OOP and its calculated measures in programming interactivity.

v.  **OOP can be used to implement interfaces which can solve real life problems:** the computed mean value = 4.45. The mean value (4.45) is far above the agreed mean value which is 3.00 this also shows that the statement is in support of OOP and its calculated measures in programming interactivity.

vi.  **Application interface interactivity is more dynamic in terms of interfaces written in OOP languages like Java, VB, C++, C#, etc:** the computed mean value = 4.08. The mean value (4.08) is above the agreed mean value which is 3.00 this shows that the statement is in support of OOP and its calculated measures in programming interactivity.

vii.  **OOP concept like polymorphism, encapsulation, inheritance, classes, abstraction helps in application development flexibility and interactivity:** the computed mean value = 4.63. The mean value (4.63) is far above the agreed mean value which is 3.00; this also shows that the statement is highly in support of OOP and its calculated measures in programming interactivity.

Comparatively, the level of OOP concepts that has helped in application development to be flexible and interactive was highest with an aggregate mean of 4.63, followed by the introduction of OOP that has made program design flexible for usage with mean of 4.58, and the implementation of interfaces with OOP to solve real life problems (4.45). The aggregate results suggest that the level of the calculated measurement of OOP were fairly high and in support since the mean values were greater than the agreed mean value of 3.00.

## V.  Conclusion

We have in this paper presented a survey on Object Oriented Programming (OOP) and its calculated measures in programming interactivity. The results of this study reveals and strongly suggested that the use of calculated measure program such as flexible design, objects design, scalability in programming, interfaces in solving real life problems, application interface dynamics, the concepts of inheritance, polymorphism, encapsulation, abstraction, etc are in support of the Object Oriented Programming (OOP) and its calculated programming interactivity. The survey reflects the importance of object concept in the interoperation of different programming languages by acting as a single entity and thereby expanding the scope and act of the current breakthrough in software development industries. This, therefore suggest that as effort and time begin to find its root in the manner at which current software designs are created, using object approach communication and interactions will become more useful to its original aim.